\documentclass[aps,prl,twocolumn,preprintnumbers,amsmath,amssymb,superscriptaddress,nolongbibliography,noeprint]{revtex4-2}
\usepackage{graphicx}
\usepackage{epsfig}
\usepackage{dcolumn}
\usepackage{bm}
\usepackage{ulem}
\usepackage{color}
\usepackage{float}
\usepackage[colorlinks=true,linkcolor=blue,citecolor=blue,urlcolor=blue]{hyperref}%
\usepackage{verbatim}
\hypersetup{citecolor = blue}
\usepackage{subfigure}
\usepackage[ruled,vlined]{algorithm2e}

\def\be{\begin{equation}}       \def\ee{\end{equation}}
\def\bea{\begin{eqnarray}}      \def\eea{\end{eqnarray}}
\def\ba{\begin{array}}
\def\ea{\end{array}}
\def\bnum{\begin{enumerate} }
\def\enum{\end{enumerate}}

\def\=>{\Rightarrow}
\def\>{\rightarrow}

\def\eye2{Fathbb{I}}

\usepackage{ listings }

\renewcommand{\>}{\rangle}

\renewcommand{\rm}[1]{\mathrm{#1}}

\usepackage{listings}
\usepackage{color}
\definecolor{lightgray}{gray}{1}

\begin{document}
\graphicspath{{figures/}}

\title{Unraveling Deconfined Quantum Criticality in non-Hermitian Easy-plane $J$-$Q$ Model}

\author{Xuan Zou}
\affiliation{Institute for Advanced Study, Tsinghua University, Beijing 100084, China}
\author{Shuai Yin}
\email{yinsh6@mail.sysu.edu.cn}
\affiliation{Guangdong Provincial Key Laboratory of Magnetoelectric Physics and Devices, School of Physics, Sun Yat-Sen University, Guangzhou 510275, China}
\affiliation{School of Physics, Sun Yat-Sen University, Guangzhou 510275, China}
\author{Zi-Xiang Li}
\email{zixiangli@iphy.ac.cn}
\affiliation{Beijing National Laboratory for Condensed Matter Physics \& Institute of Physics, Chinese Academy of Sciences, Beijing 100190, China}
\affiliation{University of Chinese Academy of Sciences, Beijing 100049, China}
\author{Hong Yao}
\email{yaohong@tsinghua.edu.cn}
\affiliation{Institute for Advanced Study, Tsinghua University, Beijing 100084, China}

\date{\today}

\begin{abstract}
Deconfined quantum critical point (DQCP) characterizes the continuous transition beyond Landau-Ginzburg-Wilson paradigm, occurring between two phases that exhibit distinct symmetry breaking. The debate over whether genuine DQCP exists in physical SU(2) spin systems or the transition is weakly first-order has persisted for many years. In this letter, we construct a non-Hermitian easy-plane $J$-$Q$ model and perform sign-problem-free quantum Monte-Carlo (QMC) simulation to explore the impact of non-Hermitian microscopic interactions on the transition that potentially features a DQCP.
Our results demonstrate that the intensity of the first-order transitions significantly diminishes with the amplification of non-Hermitian interactions, serving as numerical evidence to support the notion that the transition in $J$-$Q$ model is quasi-critical, possibly in the vicinity of the fixed point governing DQCP in the complex plane, described by a non-unitary conformal field theory (CFT). The non-Hermitian interaction facilitates the approach towards such a complex fixed point in the parameter regime.    Furthermore, our QMC study on the non-Hermitian J-Q model opens a new route to numerically investigating the nature of complex CFT in the microscopic model. 
\end{abstract}

\maketitle
{\it Introduction.}---Deconfined quantum criticality \cite{Senthil2004a,Senthil2004b,Senthil2023Review,Levin2004PRB,Motrunich2004PRB,Hu2005PRL,Senthil2006PRB,Lee2010PRB,You2018PRX,Motrunich2019PRB,Jian2018PRB,Yan2025arXiv} is a prototype scenario of continuous phase transition beyond the celebrated Landau-Ginzburg-Wilson (LGW) paradigm, tremendously renewing our understanding of phase transition. The theory of deconfined quantum critical point 
 (DQCP) is first proposed to describe   a continuous transition from the antiferromagnetic (AFM) to the valence bond solid (VBS) state \cite{Senthil2004a,Senthil2004b}, which exhibits distinct symmetry breaking—unlike the first-order transitions predicted by Landau paradigm \cite{SachdevBook}. Due to its exotic  
 physics, it has been pursued for many years through theoretical, numerical, and experimental research.  
 DQCPs have been well studied in SU(N) spin systems in large $N$ limit \cite{Senthil2004a,Senthil2004b,Kaul2012}, encapsulated by CP$^{N-1}$ field theory. 
However, the scenario becomes more intricate when considering physical SU(2) spin systems. Despite decades of research, the definitive demonstration of DQCPs in these systems remains a vastly debated topic. Recent experimental results reported evidence of proximate deconfined quantum criticality in the compound SrCu$_2$(BO$_3$)$_2$ \cite{cui2023proximate,cui2025,Cui_2025}, while other studies advocate for a first-order transition in this compound~\cite{guo2023deconfined}, thus stimulating ongoing discussions on this topic, including the effects of spin–lattice coupling \cite{Hofmeier2024}.

\begin{figure}[htbp]
\centering
\includegraphics[width=0.8 \columnwidth]{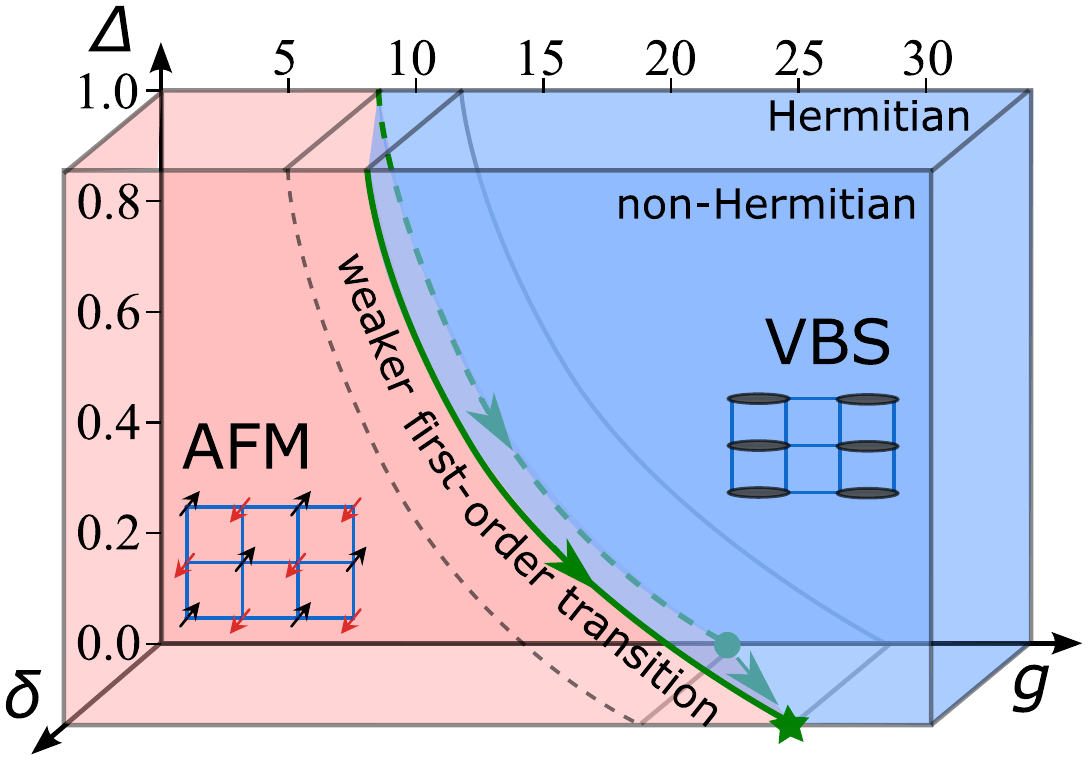}
\caption{  A sketch of the phase diagram of the NHJQ model is presented, depicting the transition lines with (solid green line) and without (dashed green line) non-Hermitian interaction. Here, \(\Delta\) represents the strength of the easy-plane term, \(\delta\) signifies the strength of non-Hermitian interaction, and \(g = \frac{Q}{J}\) corresponds to the ratio between the strengths of two interaction terms in J-Q model. The green dot marks the transition point in the original $J$-$Q$ model. For clarity and comparison, the transition line from the Hermitian case \(\delta = 0\) is also presented in the non-Hermitian \(\delta > 0\) regime, and vice versa. The arrows indicate that the strength of the first-order transition becomes weaker. The green star denotes the transition point under non-Hermitian interaction at SU(2) limit, signifying the weakest first-order transition in the phase diagram, which potentially could be interpreted as a proximate continuous transition.}
\label{fig1}
\end{figure}

Due to the theoretical complexities inherent to the CP$^1$ theory, numerical methodologies are indispensable in elucidating the intricate phenomena of DQCP. 
An early microscopic platform is the Shastry-Sutherland model (SSM), relevant to SrCu$_2$(BO$_3$)$_2$ \cite{SS1981,Koga2000,sstensor,Wessel2018,Wietek2019,Lee2019,Yang2022,Wang_2022,NXi2023,liu2024,chen2025}. 
Another widely used SU(2) model featuring a Néel to VBS phase transition is the $J$-$Q$ model~\cite{sandvik2007,Kaul2008prl,Troyer2008PRL,Sandvik2011PRL,Kaul2013Review,Chen2013PRB}, along with its generalizations~\cite{Lou2009,Sandvik2012,BWZhao2020,Ribhu2016,Ribhu2020,Xiang2012PRB,Gu2022PRX,Alet2015PRB,Zhang2018PRL,Wang2016PRB}. Additionally, a multitude of models have been scrutinized as viable platform for the manifestation of DQCPs, including loop models~\cite{Nahum2015,sreejithEmergent2019}, an array of fermionic systems~\cite{gAZIT2017NP,Grover2016prx,li2017fermion,Sato2017,HongYao2019,liu2019superconductivity,ZJWang2021a,ZJWang2021b,Liao2022a,Assaad2023PRB,liu2022fermion,YHLiu2022,ZXL2024PRL,wu2024,liu2025,SongXY2024,yu2025,myersonjain2024}.
The early investigations into the $J$-$Q$ model imply the nature of continuous phase transition \cite{sandvik2007}, even in the large cluster with system size up to tens of thousands of spins \cite{sandvik2010}. Nevertheless, the critical exponents display a pronounced dependence on the size of the simulated systems, distinct from conventional continuous transitions.
Moreover, with the advent of increasingly refined results from conformal bootstrap, the critical exponents discerned within the $J$-$Q$ model, among others, have been identified to contravene the established conformal bounds \cite{Nakayama2016,Poland2019}.
The potential strategies to address these challenges encompass various hypotheses, including the two-length scale scaling \cite{Shao2016,Yin2022PRL}, multi-critical points \cite{HongYao2019,BWZhao2020,DCLu2021,sandvik2024,Chen2024,Sandvik2024arXiv,Su2024PRL,Yin2025nc}, and pseudo-criticality \cite{Nahum2015,CWang2017,RCMa2020,Nahum2020}. 

The prevailing hypothesis posits that the phase transitions in J-Q model and other microscopic models possibly manifest as an extremely weak first-order transition. Intriguingly, several notable properties, such as enhanced symmetry \cite{BWZhao2020,Zhao2019NP,Yu2023PRB} and dynamical signature of fractionalization \cite{NSMa2018}, have been observed. A viable explanation for these phenomena posits that the associated fixed points reside on the complex plane, closely proximate to the real axis. Consequently, for finite systems, the transition behavior closely resembles that of a continuous transition—this is characterized as pseudo-critical behavior.
Furthermore, these complex fixed points are believed to be characterized by a non-unitary conformal field theory (CFT). 
Moreover, possible numerical evidence that the AFM–VBS transition is governed by a non-unitary CFT or exhibits pseudo-criticality has been reported using the newly developed fuzzy-sphere framework \cite{Zhu2023PRX,Zhu2023PRL,Zhu2023arXiv,yang2025}.
Nonetheless, the direct numerical evidence for the existence of complex fixed point governing the DQCP in a microscopic model is rare, partially due to the scarcity of the numerically solvable non-Hermitian microscopic model hosting phase transition potentially described by non-unitary CFT.  
This raises a compelling inquiry: is it possible to construct an intrinsically non-Hermitian microscopic model potentially featuring the non-unitary CFT at low energy and numerically solve the model through unbiased approach, hence providing convincing evidence for the scenario of DQCP governed by complex fixed point?


To shed the light on the complex fixed points corresponding to DQCP, we propose the non-Hermitian $J$-$Q$ model, and investigate the impact of the non-Hermitian interactions on the phase transition properties. Remarkably, the model is sign-problem-free in quantum Monte-Carlo (QMC) simulation, enabling numerically exact QMC simulation on the model with large system size and low temperature \cite{Sandvik2010Review,Sugar1990PRB,ZXLi2015PRB,ZXLi2016PRL,Yao2019Review,Yu2024arXiv}. To our knowledge, it is the first QMC study on the non-Hermitian quantum spin model in (2+1)-dimension. 
To make the strength of the first order transition tunable, we include the easy-plane interaction $\Delta$.
As shown in Fig. \ref{fig1}, 
the non-Hermitian interactions $\delta$ appears to shift the transition points.
By scrutinizing the discontinuity of the order parameters at the transition points, we identify that the non-Hermitian interactions notably diminish the strength of first-order transitions. This is further supported by the behavior of the critical exponents.
Our results suggest that non-Hermitian interactions facilitate the approach toward continuous transition points, providing numerical signature supporting the notion that the fixed point associated with DQCP is located in the complex plane.

{\it Model and Methods.}---To systematically investigate the strength of first-order phase transitions, we consider the non-Hermitian easy-plane $J$-$Q$ (NHJQ) model:
\begin{equation}
H = (1 - \Delta) H_{\mathrm{JQ}} + \Delta H_{\mathrm{ep}} + \delta H_{\mathrm{nH}},
\label{eq_H}
\end{equation}
where \( \Delta \) parameterizes the interpolation between the standard $J$-$Q$ model \( H_{\mathrm{JQ}} \) and the easy-plane $J$-$Q$ model \( H_{\mathrm{ep}} \). The last term, \( H_{\mathrm{nH}} \), represents the non-Hermitian term, with \( \delta \) indicating the strength of the non-Hermitian interaction.

The Hamiltonian of the original $J$-$Q$ model is given by \cite{sandvik2007}
\begin{equation}
    H_{\mathrm{JQ}}(\{\mathbf{P}_{ij}\}) = -J \sum_{\langle i, j \rangle} \mathbf{P}_{ij} - Q \sum_{\langle i, j, k, l \rangle} \mathbf{P}_{ij} \mathbf{P}_{kl},
\label{eq_su2JQ}
\end{equation}
where $\mathbf{P}_{ij} = \frac{1}{4} - \mathbf{S}_i \cdot \mathbf{S}_j$ represents the singlet projection operator for sites $i$ and $j$. Here, the $J$ term corresponds to the Heisenberg interaction, with $\langle i, j \rangle$ indicating summation over nearest-neighbor pairs. Conversely, the $Q$ term introduces a four-spin interaction, with $\langle i, j, k, l \rangle$ denoting the sites forming a plaquette.
For the easy-plane variant of the $J$-$Q$ model \cite{Ribhu2020}, denoted as $H_{\mathrm{ep}}(\{\tilde{\mathbf{P}}_{ij}\})$, the modified projection operator $\tilde{\mathbf{P}}_{ij} = S_i^x S_j^x + S_i^y S_j^y$ is used to encapsulate the easy-plane anisotropy within the $J$-$Q$ framework.

The non-Hermitian term is introduced as follows: 
\begin{equation}
H_{\text{nH}} =\sum_{\langle i, j \rangle} \frac{J}{2} S_i^{+} S_j^{-}  -\frac{J}{2} S_i^{-} S_j^{+},
\end{equation}
where the relative strength of the non-Hermitian interaction is quantified by $\delta$ in Eq. \ref{eq_H}, varying from 0 to 1.  
Here, $\langle i, j \rangle$ denotes the pair of nearest neighbor sites, with $i$ and $j$ specifically positioned as follows: for the $x$-bond, $i$ is on the left and $j$ is on the right; for the $y$-bond, $i$ is at the bottom and $j$ is at the top.

To investigate the phase diagram and the nature of the phase transitions, we apply the stochastic series expansion (SSE) quantum Monte Carlo (QMC) method \cite{Sandvik1991,Sandvik1992}. Adapting the SSE method to the non-Hermitian case is straightforward \cite{Shixin2023,Zhang2025PRL} and NHJQ is a sign-free model as long as the strength of non-Hermitian interaction $|\delta|<1$\cite{Sandvik2010Review,Li2024PRL}. 
For the standard SU(2) symmetric $J$-$Q$ model with \( \Delta = 0 \), the strength of the first-order transition is extremely weak and difficult to discern in simulations. In order to more clearly observe the effects of non-Hermitian interactions on the discontinuity of the transition, our investigations primarily focus on the case of \( \Delta = 0.6 \). The simulation results for \( \Delta = 0 \) are discussed subsequently.

To ascertain the phase transition point and evaluate the strength of the first-order transition in the NHJQ model, we consider the in-plane AFM and VBS orderings. Specifically, the AFM order parameter is defined as $m_a=\frac{1}{N} \sum_i(-1)^{\left(x_i+y_i\right)} S^x(i)$  and the VBS order parameter is defined as $m_v=\frac{1}{N} \sum_i(-1)^{x_i} Q^x(i)$. Here, $S^x(i)$ represents the $x$-direction spin operator on site $i$ and \(Q^x(i)\) represents the plaquette operator\cite{Ribhu2020}, with the details included in SM. Following these definitions, we compute the structure factor and the RG-invariant correlation ratio to analyze the phase transitions.

\begin{figure}
\centering
\includegraphics[width=1.0\columnwidth]{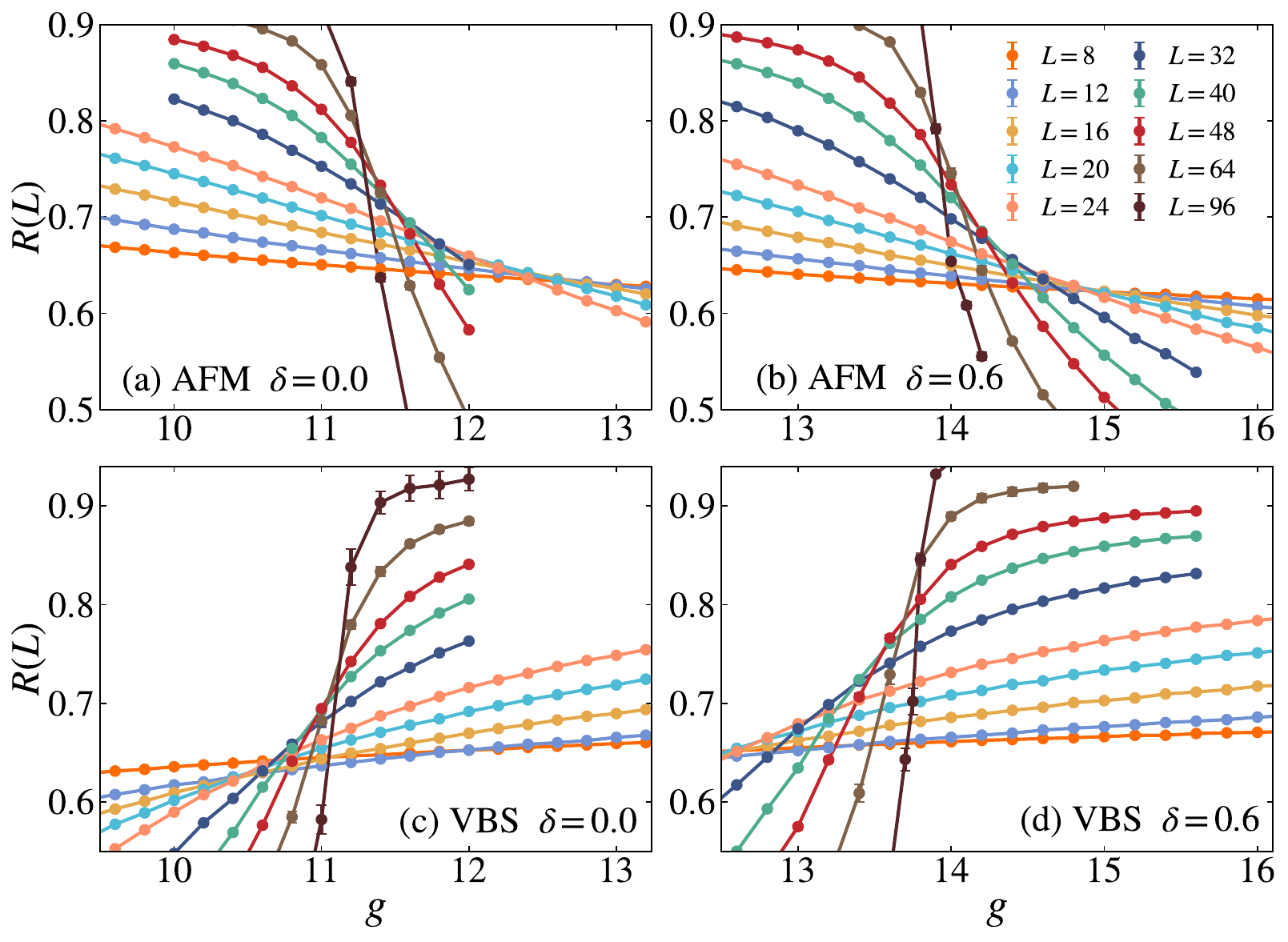}
\caption{ At $\Delta=0.6$, the RG-invariant correlation ratios for the  AFM order $ R(L) $ vary as a function of the coupling ratio $ g = Q/J $ for different system sizes, under different strengths of non-Hermitian coupling:  (a) $\delta = 0.0$ and (b) $ \delta = 0.6 $, respectively. 
The intersection points in these plots demarcate the phase transition points. Panels (c) and (d) display the RG-invariant correlation ratios for the VBS order. The phase transition points from AFM to VBS states shift towards higher values of $g$ with an increase in the strength of the non-Hermitian coupling parameter $\delta$.}
\label{fig2}
\end{figure}

The spin structure factor of the order under consideration is defined as 
\begin{equation}
S(\vec{q}) = \frac{1}{N} \sum_{i, j} \langle \hat{O}(i) \hat{O}(j) \rangle e^{i \vec{q} \cdot (\vec{r}_{i} - \vec{r}_{j})}.
\end{equation}
For in-plane AFM order $\hat{O}(i)$ = $S^x(i)$ and the peak of momentum is expected at \( \vec{q}^* = (\pi, \pi) \). For VBS order $\hat{O}(i)$ = $Q^x(i)$ with the characteristic peak of momentum expected at \( \vec{q}^* = (\pi, 0) \).

The RG-invariant correlation ratio is defined as 
\begin{equation}
R = 1 - \frac{S(\vec{q}^* + \delta \vec{q})}{S(\vec{q}^*)}, 
\end{equation}
where $\vec{q}^*$ is the peaked momentum of the associated order parameter and $\delta \vec{q} = (\frac{2\pi}{L},\frac{2\pi}{L})$ is a minimum momentum on lattice. The value of \( R \) trends towards one in the ordered phase and towards zero in the disordered phase as the system size increases. The phase transition point is inferred from the crossing points of the correlation ratios for different system sizes.

{\it Numerical Results.}---Before presenting the details of the numerical results, we summarize the main features of the ground-state phase diagram as illustrated in Fig.~\ref{fig1}. The parameter \( \Delta \) characterize the anisotropy of spin interaction, with the conventional $J$-$Q$ model defined at \(\Delta = 0.0\) and the pure easy-plane scenario at \(\Delta = 1.0\). The variable \( g \) represents the coupling ratio, formally defined as \( Q/J \) in Eq.~\ref{eq_su2JQ}, whereas \(\delta\) denotes the strength of the non-Hermitian interaction.
Within the \(\Delta\)-\(g\) plane (at \(\delta = 0\)), the AFM-VBS transition line for the Hermitian easy-plane $J$-$Q$ model is delineated by a green dashed line. The critical point in the standard $J$-$Q$ model, indicative of a weak first-order transition, is highlighted by a green dot.
In cases where \(\delta > 0\), signifying the presence of non-Hermitian interactions, the phase transition is depicted by a green solid line on the \(\Delta\)-\(g\) plane. The green arrows indicate that the strength of the first-order transition becomes weaker. The transition point under non-Hermitian interaction, marked by a green star, represents continuous transition or a weaker first-order transition compared to the Hermitian case. Here, the phase boundary between AFM and VBS phases is determined by the RG-invariant correlation ratio of corresponding order parameters. The discontinuity of the order parameters crossing the transition point serves as the indicator of the strength of first-order transitions. 

In the standard $J$-$Q$ model with \( \Delta = 0 \), the first-order transition is notably weak and challenging to identify, making it difficult to detect and analyze the effects of added non-Hermitian interactions.  
Therefore, we primarily focus our study on the easy-plane scenario, characterized by \( \Delta>0\), where the signs of the first-order transition are more evident.
At \( \Delta = 0.6 \), Fig. \ref{fig2} (a-b) illustrates that the RG-invariant correlation ratio \( R(L) \) for the in-plane AFM order tends towards 1 for \( g < g_c \) but drops to 0 for \( g > g_c \) as the system size increases. 
In contrast, for the VBS order, illustrated in (c-d), the behavior of \( R(L) \) is reversed.
The crossing points of the \( R(L) \) curves serve as indicators of the phase transition points, exhibiting consistent results across both AFM and VBS order parameters. Moreover, it is observed that increasing the non-Hermitian coupling strength, \( \delta \), results in a shift of the transition points to higher \( g_c \) values. (Detailed results for varying \( \delta \) levels are available in the Supplementary Material.) Previous investigations into the hermitian $J$-$Q$ model revealed that as the strength of the easy-plane decreases, the intensity of the first-order transition diminishes, while concurrently, the critical points, denoted as $g_c$, shift to higher values. 

\begin{figure}
\centering
\includegraphics[width=1.0\columnwidth]{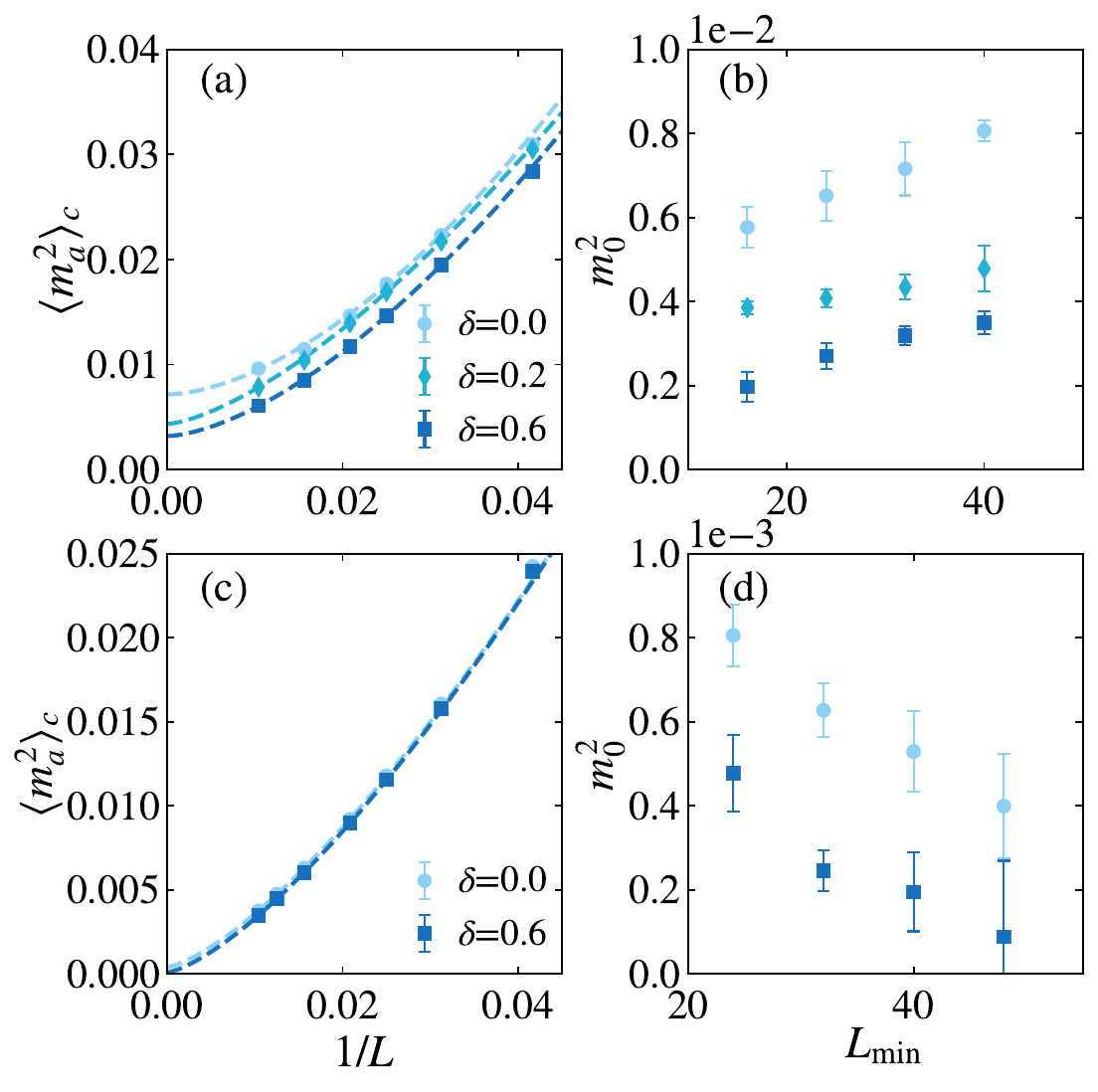}
\caption{ 
(a,c) The order parameters for the AFM order are examined as functions of the system size at the critical transition point, for various strengths of non-Hermitian interactions at (a) $\Delta=0.6$ and at (c) $\Delta=0.0$.
The analysis in (a) and (c) employs fitting curves based on the power-law function as $m_a^2(L)=m^2_0+b/L^c$. (b,d) The fitting results $m_0^2$ of order parameters with a series of minimum system sizes \( L_{\text{min}}\) utilizing at (b) $\Delta=0.6$ and at (d) $\Delta=0.0$. The discontinuities $m_0^2$ in the order parameters are interpreted as indicators of the strength of first-order phase transitions.
A decrease in these discontinuities suggests that the first-order transitions become less pronounced as the strength of the non-Hermitian interaction increases. }
\label{fig3}
\end{figure}
To accurately evaluate the intensity of the first-order transition, it is critical to determine the abrupt changes in the order parameters at the critical points.
Initially, we identify the intersection points between the RG-invariant correlation ratios for system sizes $L$ and $L/2$, thus determining the finite-size transition point, denoted as $g_c(L)$. Subsequently, we ascertain the values of the order parameter, $\langle m^2 \rangle_c$, at these transition points $g_c$ and plot them against the system size to demonstrate their scaling behavior, as illustrated in Fig.~\ref{fig3}. For data analysis, we employ a power-law fitting approach using the formula $m^2(L) = m^2_0 + bL^{-c}$, which renders the extrapolation of  $m_0^2$, namely the square of order parameter in the thermodynamic limit.

\begin{figure}
\centering
\includegraphics[width=1.0\columnwidth]{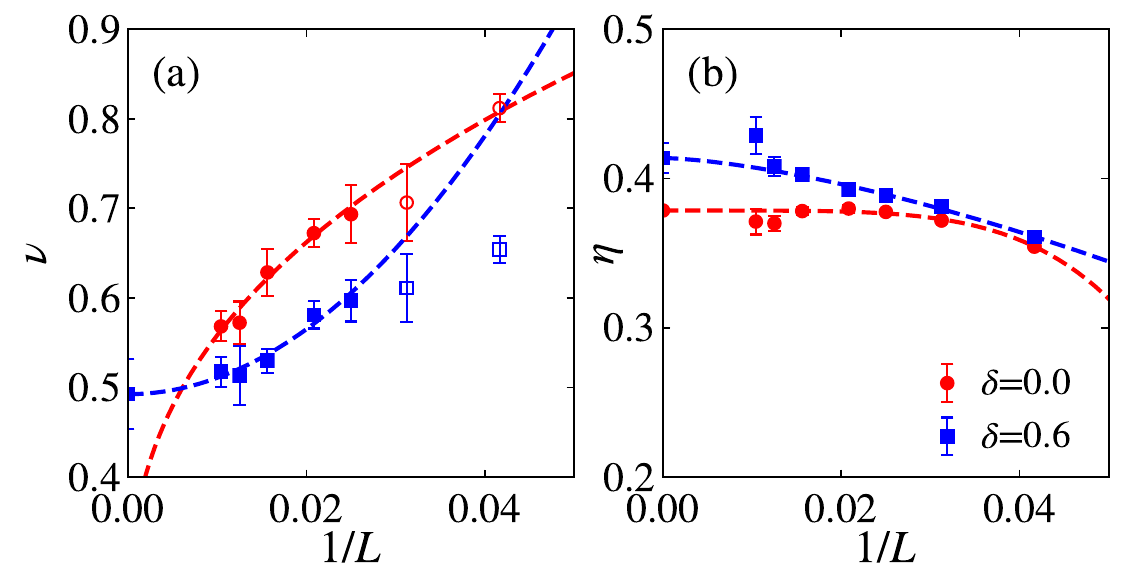}
\caption{ At $\Delta=0.0$,
the critical exponent (a)  $\nu$ and (b) $\eta$ are examined as functions of the system size at the critical transition point  for both the Hermitian case (red) and the non-Hermitian case (blue).
(a) The analysis utilizes fitting curves based on the power-law function as $\nu(L)=\nu+b/L^c$, employing a minimum system size of $L_{\text{min}}=40$. The empty data points are excluded from the fitting.  The fitting results is $\nu_0(\delta=0.6)\approx0.49(4)$.
(b) The analysis utilizes fitting curves based on the power-law function as $\eta(L)=\eta+b/L^c$, employing a minimum system size of $L_{\text{min}}=24$. The fitting results are $\eta_0(\delta=0.0)\approx0.38(1)$ and $\eta_0(\delta=0.6)\approx0.41(1)$.}
\label{fig4}
\end{figure}

Fig.~\ref{fig3} displays the discontinuities in the AFM order parameters at the transition points for different system sizes at (a-b) $\Delta=0.6$ and at (c-d) $\Delta=0.0$. The findings reveal that the magnitude of the first-order transition is influenced by the strength of the non-Hermitian coupling.
This decreasing trend is less evident for the VBS order and the fitting results shifts with different minimum sizes utilized (see SM). Previous research suggests the necessity of employing larger system sizes for the VBS order to achieve reliable fitting results that accurately reflect the thermodynamic limit \cite{Ribhu2020}. 
Additionally, the more noticeable shift in critical points for the VBS order, as compared to the AFM order, corroborates this observation (see SM). The fitting outcomes and corresponding illustrations, showing their dependency on various minimum system sizes, are detailed in the SM.

At \(\Delta = 0.6\), as depicted in Fig.~\ref{fig3}(a), the order parameters at the critical points are examined as functions of system size, with the fitting curves, based on the power-law function as $m_a^2(L)=m^2_0+b/L^c$. Fig.~\ref{fig3}(b) illustrates the fitting outcomes, specifically $m_0^2$, as extrapolated to the thermodynamic limit, considering different minimum system sizes employed in the fitting process. 
The discontinuities $m_0^2$ in the order parameters are interpreted as indicators of the strength of first-order phase transitions.
Although the values of $m_0^2$ shift with different minimum sizes $L_{\text{min}}$ used, as shown in Fig.~\ref{fig3} (b), the trend that the first-order transition weakens as $\delta$ increases remains consistent.
At \(\Delta = 0.0\), utilizing the same procedure of analysis, this trend remains robust.
Significantly, the order parameter approaches zero within fitting error for \(L_{\text{min}} = 48\) and \(\delta = 0.6\), as shown in Fig.~\ref{fig3}(d). 

Fig.~\ref{fig4} depicts (a) the correlation-length exponent $\nu$ and (b) the exponent of anomalous dimension $\eta$ with employing different minimum system sizes $L_{\text{min}}$. The methods of extracting the critical exponents are presented in SM. 
The exponent $\nu$ at $\delta=0.6$ converges as $L$ increases. This is in sharp contrast to the Hermitian case wherein the remarkable drift of $\nu$ on size is usually regarded as the signature of an ultimate first-order phase transition. Therefore, the fact that $\nu$ tends to a stable value with non-Hermitian hopping Hamiltonian indicates that the strength of discontinuity is enormously weakened in the presence of non-Hermitian interaction. The transition is continuous or an extremely weakly first-order transition in vicinity of a complex fixed point.  

Moreover, the fitting results with employing a minimum system size of $L_{\text{min}}=24$ are $\eta(\delta=0.0)\approx0.38(1)$ and $\eta(\delta=0.6)\approx0.41(1)$, as shown in  Fig.~\ref{fig4} (b).  The fitting results with $L_{\text{min}}=24$  for the non-Hermitian case of $\eta$ at $\delta=0$ is consistent with the results $0.35(3)$ in Ref. \cite{Melko2008} but larger than $0.26(3)$\cite{sandvik2007} and $0.27(1)$ \cite{Sandvik2012}. Such deviation is possible due to the different procedures of the fitting (see SM). Hence, the critical exponent $\eta$ is larger in the NHJQ model, compared with the Hermitian counterpart.  The increment of $\eta$ in non-Hermitian Hamiltonian also reinforces the hypothesis that non-Hermitian interactions reduce the intensity of the first-order transition, potentially resulting in a continuous transition governed by the fixed points residing in a complex parameter space.

{\it Discussions and Conclusions.}---In fact, the non-unitary CFT scenario has also been applied to describe the first-order transition of the \(q\)-Potts model with \(q > 4\), known as the walking mechanism \cite{Gorbenko_2018}. The weakly first-order transition with the fixed point located on the complex plane has been proven in \cite{Haldar_2023}. Remarkably, a recent study on the non-Hermitian quantum 5-Potts model has identified two complex conjugate critical points and achieved a continuous phase transition \cite{Zhu2024PRL}. Our sign-problem-free QMC study on the NHJQ model paves a new route to constructing non-Hermitian microscopic model and investigating the non-unitary CFT through approximation-free numerical approach.  In the context of DQCP, it 
is straightforward to generalize our NHJQ model to other cases, for example SU(N) spin, to identify and explore the features of possible deconfined criticality described by the non-unitary CFT, which is left for our future work.


In conclusion, our study delves into the intriguing AFM-VBS phase transition in the easy-plane $J$-$Q$ model under the influence of non-Hermitian interactions. Our results of sign-problem-free QMC simulation demonstrate that with the amplification of non-Hermitian interaction strength, there is a noticeable shift in the critical points towards higher values, accompanied by a reduction in the intensity of the first-order transition in both the easy-plane and isotropic $J$-$Q$ models. Although we cannot exclude the possibility that the AFM-VBS transitions featured in NHJQ model is still weakly first-order, the inclusion of Hermitian interactions in the microscopic model tremendously diminish the discontinuity of the phase transition, supporting that the fixed point governing the deconfined criticality is possibly situated on a complex plane. Our construction of non-Hermitian quantum spin model offers a novel pathway towards realizing a genuine continuous transition featuring deconfined quantum criticality in microscopic models.

{\it Acknowledgements}: We would like to thank Yu-Min Hu, Zhou-Quan Wan, and Wei Zhu for helpful discussions. This work is supported in part by the NSFC under Grant Nos. 12347107 (X.Z., Z.X.L. and H.Y.), 12334003 (X.Z. and H.Y.) and 12474146 (Z.X.L), MOSTC under Grant No. 2021YFA1400100 (H.Y.), the New Cornerstone Science Foundation through the Xplorer Prize (H.Y.) and Beijing Natural Science Foundation under Grant No. JR25007 (Z.X.L.). S.Y. is supported by the National Natural Science Foundation of China (Grants No. 12222515), the Research Center for Magnetoelectric Physics of Guangdong Province (Grants 2024B0303390001), and the Guangdong Provincial Key Laboratory of Magnetoelectric Physics and Devices (Grant No. 2022B1212010008). S.Y. is also supported by the Science and Technology Projects in Guangdong Province (Grant No. 2021QN02X561) and Guangzhou City (Grant No. 2025A04J5408).

\bibliography{ref}

\newpage
\begin{widetext}
\section*{Supplemental Materials}
\renewcommand{\theequation}{S\arabic{equation}}
\setcounter{equation}{0}
\renewcommand{\thefigure}{S\arabic{figure}}
\setcounter{figure}{0}
\renewcommand{\thetable}{S\arabic{table}}
\setcounter{table}{0}

In the Supplemental Material, we present further numerical results obtained for the easy-plane cases $\Delta = 0.6$ and $\Delta = 0$.

\subsection{Numerical results at $\Delta=0.6$}
To explore the effect of non-Hermitian coupling strengths on the critical points at $\Delta=0.6$, Fig. \ref{figs1} shows the RG-invariant correlation ratios for the AFM and VBS orders. These quantities are examined for several coupling strengths $\delta = 0.0, 0.2, 0.6,$ and $1.0$. The intersection points of the correlation-ratio curves mark the locations of the phase transitions. It is observed that the phase transition points progressively shift towards higher values as the strength of the non-Hermitian coupling parameter $\delta$ is increased. Comparative analyses with previous investigations into the Hermitian $J$-$Q$ model have shown that a decrease in the strength of the easy-plane parameter leads to a reduction in the intensity of the first-order transition. Concurrently, the critical points, denoted as $g_c$, migrate to higher values. This trend suggests that an upward shift in $g_c$ values could potentially indicate a weakening in the first-order transition.

In Fig. \ref{figs2}, we explore the values of the order parameters at the critical point versus system size.
Our analysis employs fitting curves based on the power-law function $m^2(L) = m^2_0 + b/L^c$, utilizing various minimum system sizes from $L_{\text{min}} = 16$ to 40. Panels (a-d) reveal that the extrapolated values of the AFM order parameter at the critical point, $m_0^2$, diminish as the non-Hermitian coupling increases, independent of $L_{\text{min}}$. Conversely, the fitting outcomes for the VBS order parameter demonstrate a noticeable shift with different $L_{\text{min}}$ values, suggesting that larger system sizes may be required to obtain more reliable results. 
Moreover, the magnitude of the VBS order is an order of magnitude smaller than that of the AFM order, indicating potentially more challenging computations. 
Drawing from the more reliable results of the AFM order parameter, we deduce that non-Hermitian interactions tend to mitigate the intensity of first-order transitions as the strength of non-Hermitian coupling escalates.

\begin{figure}[h]
\centering
\includegraphics[width=0.8\columnwidth]{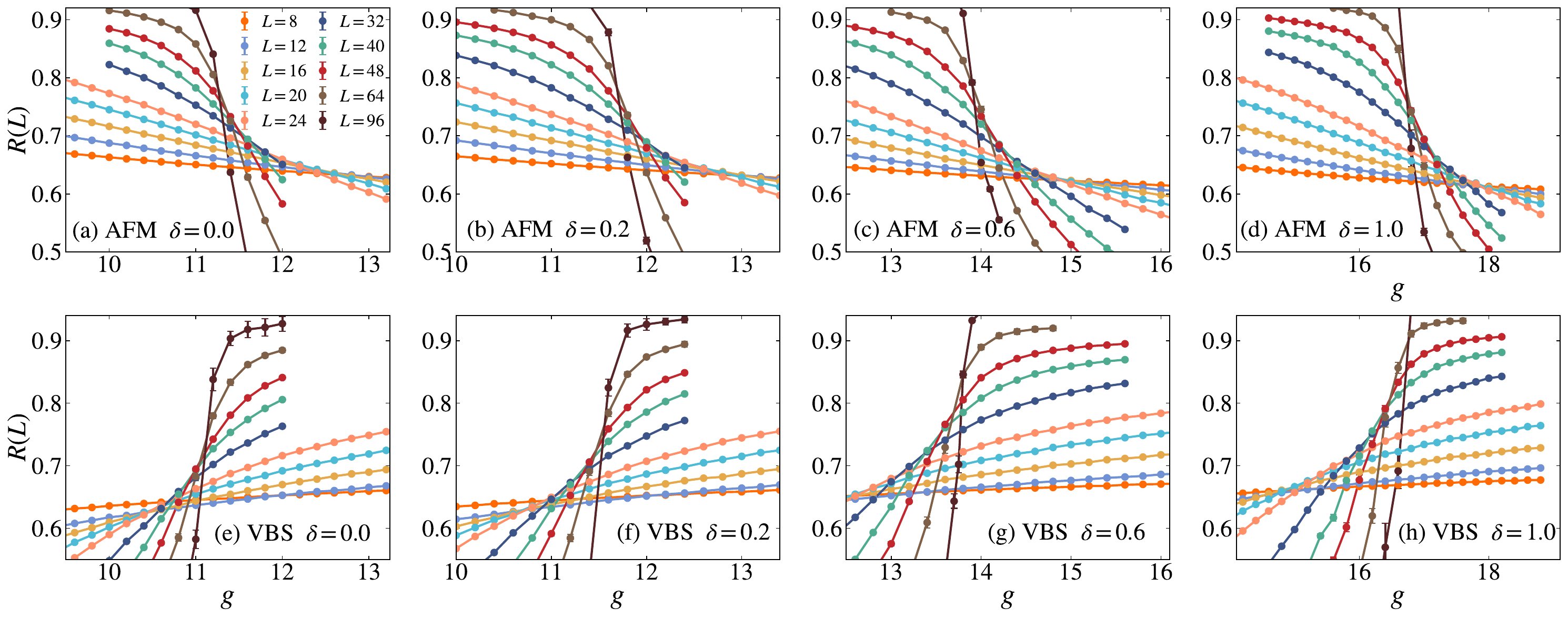}
\caption{ At $\Delta=0.6$, the RG-invariant correlation ratio for (a-d) the AFM order and (e-h) the VBS order, $ R(L) $, measured across various system sizes, vary as a function of the coupling ratio $ g = Q/J $, under different strengths of non-Hermitian coupling $\delta=0.0,0.2,0.6$ and 1.0. The intersection points in these plots demarcate the phase transition points. The phase transition points shift towards higher values with an increase in the strength of the non-Hermitian coupling parameter $\delta$.}
\label{figs1}
\end{figure}

\begin{figure}[h]
\centering
\includegraphics[width=0.8\columnwidth]{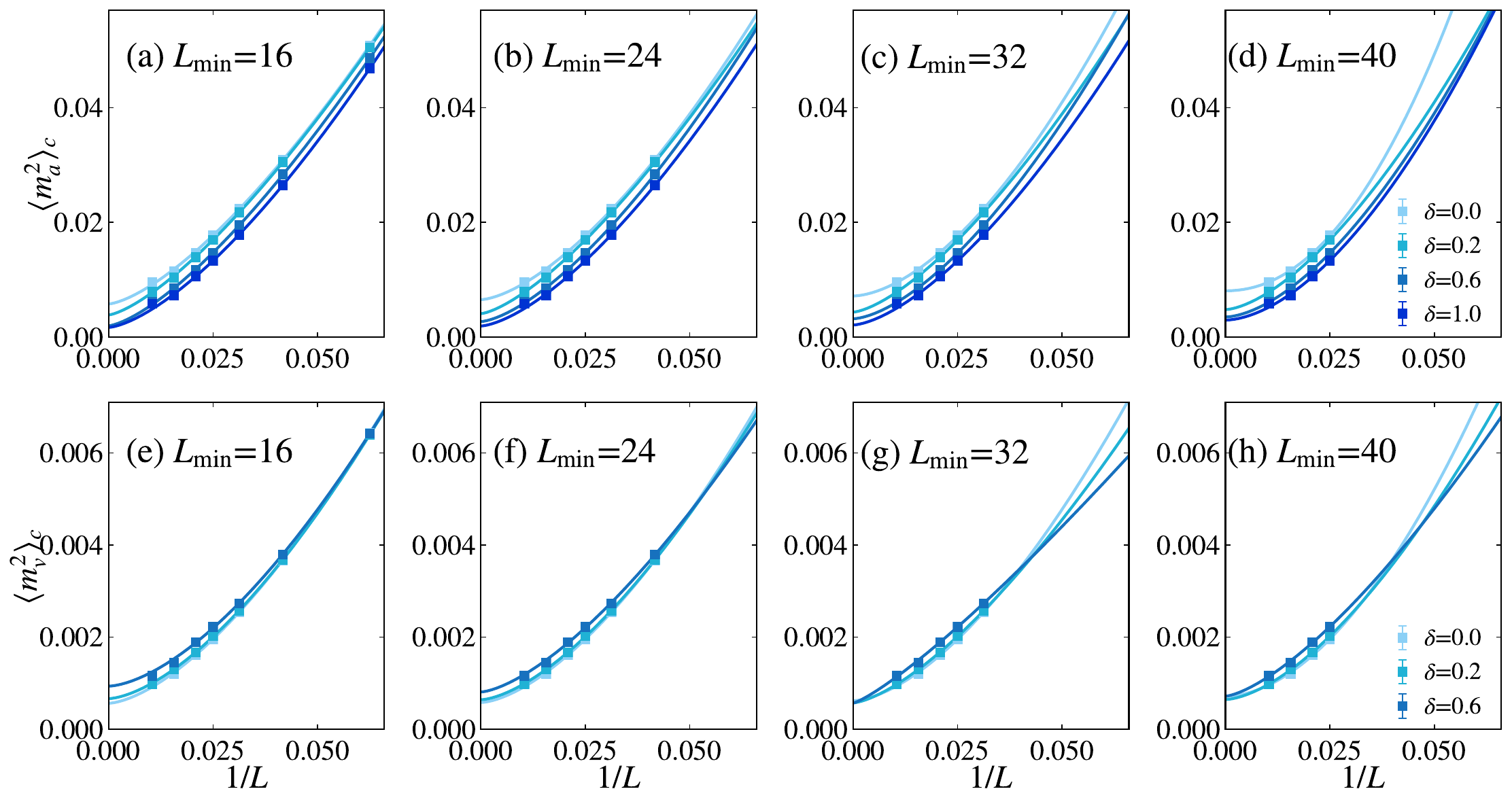}
\caption{ At $\Delta=0.6$, the order parameters for (a-d) the AFM order and (e-h) the VBS order are examined as functions of the system size at the critical transition point, for various strengths of non-Hermitian interactions. The discontinuities in the order parameters are interpreted as indicators of the strength of first-order phase transitions. 
The analysis utilizes fitting curves based on the power-law function as $m^2(L)=m^2_0+b/L^c$, with employing different minimum system size $L_{\text{min}}=16, 24, 32$ and 40.}
\label{figs2}
\end{figure}

\subsection{Numerical results at $\Delta=0$}
In an exploration of the impact of non-Hermitian coupling strengths on the critical points at $\Delta=0$,  Fig. \ref{figs3} illustrates the RG-invariant correlation ratios for the AFM and VBS orders for non-Hermitian coupling strengths $\delta = 0.0$ and $0.6$. 
It is observed that the phase transition points tend to migrate towards higher values as the strength of the non-Hermitian coupling parameter $\delta$ increases.

The shift of critical points across different system sizes is captured in Fig. \ref{figs4}. 
These critical points are identified through the intersection points of the RG-invariant correlation ratios, $R(L)$ and $R(L/2)$. Consistent with our discussion in main text, the shift in the critical points in the VBS case appears more pronounced than in the AFM case. This suggests that larger system sizes might be necessary to achieve more reliable results for the VBS order compared to the AFM order.

Delving into the correlation length exponent $\nu$, Fig. \ref{figs5} depicts its values with employing different minimum system size $L_{\text{min}}=16, 24, 32$ and 40.  The exponent $\nu$  exhibits a reduced dependence on the system size and converges to a larger value in the non-Hermitian scenario compared to the Hermitian case. 
This finding underscores the distinct impact of non-Hermitian coupling on the critical behavior, suggesting that non-Hermitian interactions may alter the nature of phase transitions.
\begin{figure}[h]
\centering
\includegraphics[width=0.8\columnwidth]{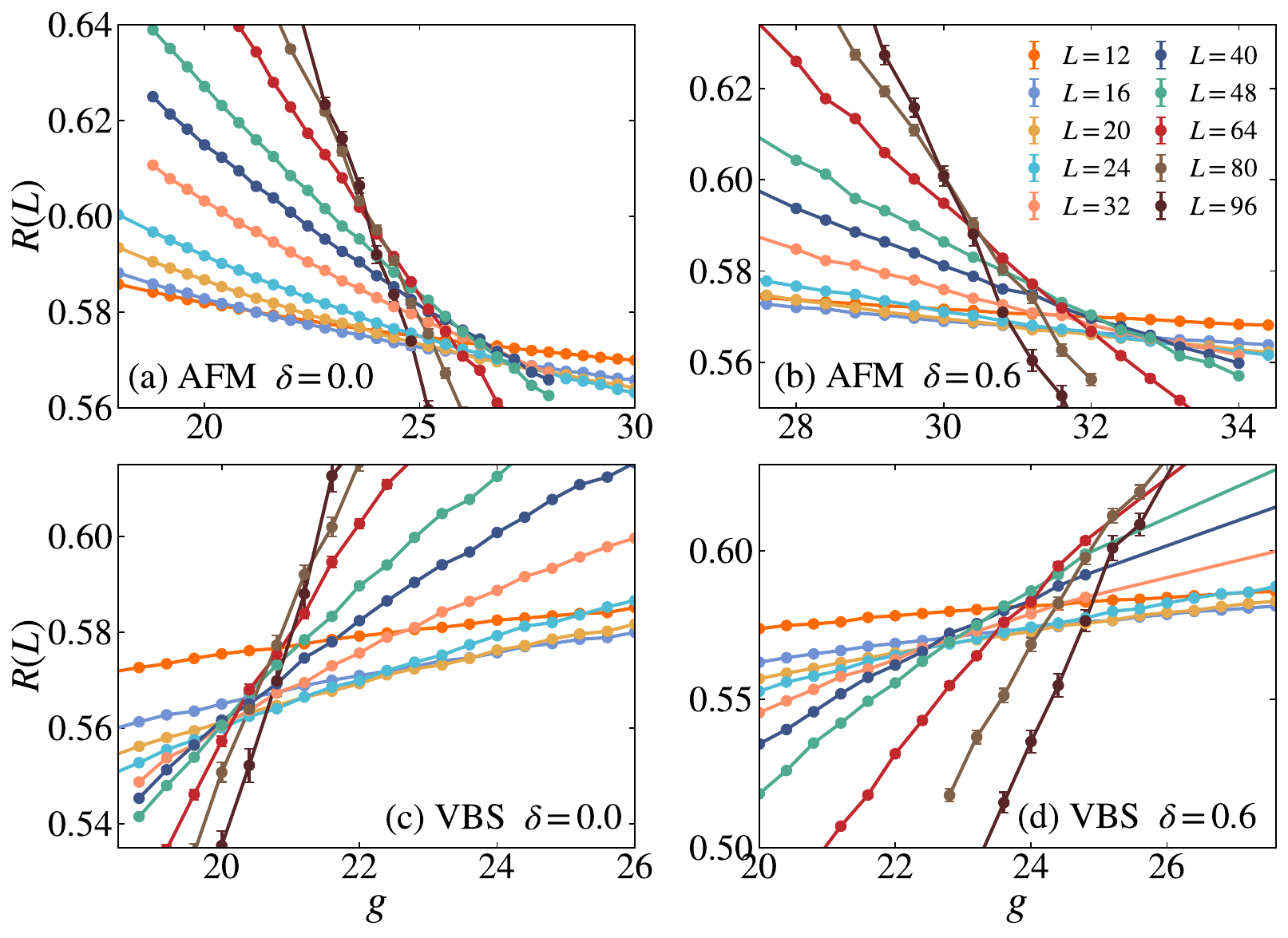}
\caption{ The RG-invariant correlation ratio at $\Delta=0.0$, $ R(L) $, measured across various system sizes, vary as a function of the coupling ratio $ g = Q/J $, under different strengths of non-Hermitian coupling:  (a,c) $\delta = 0.0$ and (b,d) $ \delta = 0.6 $, respectively. The intersection points in these plots demarcate the phase transition points. The phase transition points shift towards higher values with an increase in the strength of the non-Hermitian coupling parameter $\delta$.}
\label{figs3}
\end{figure}

\begin{figure}[h]
\centering
\includegraphics[width=0.8\columnwidth]{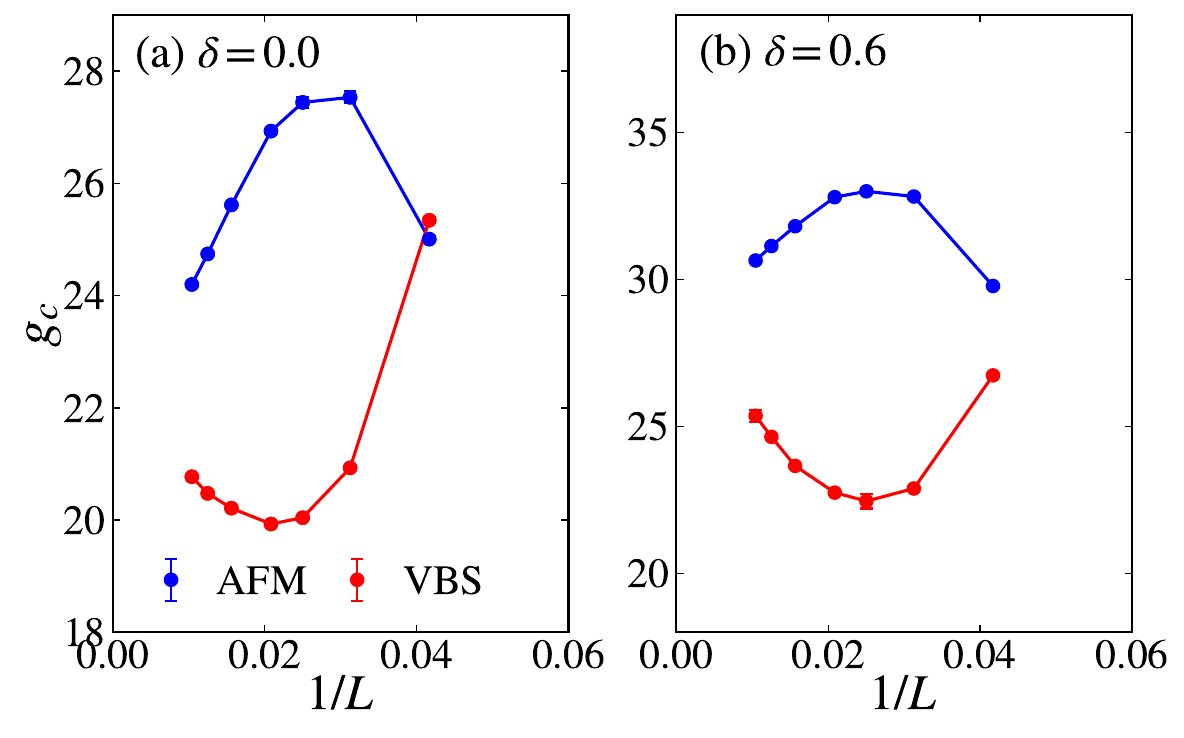}
\caption{ The critical points \(g_c(L)\) are determined through the intersection of \(R(L)\) and \(R(L/2)\) at \(\Delta = 0.0\). The shift in critical points for the VBS order (red line) is more discernible compared to the AFM order (blue line). Moreover, the variation in critical points under (b) non-Hermitian conditions with $\delta=0.6$ is less pronounced than (a) Hermitian case with $\delta=0$.}
\label{figs4}
\end{figure}

\begin{figure}[h]
\centering
\includegraphics[width=0.9\columnwidth]{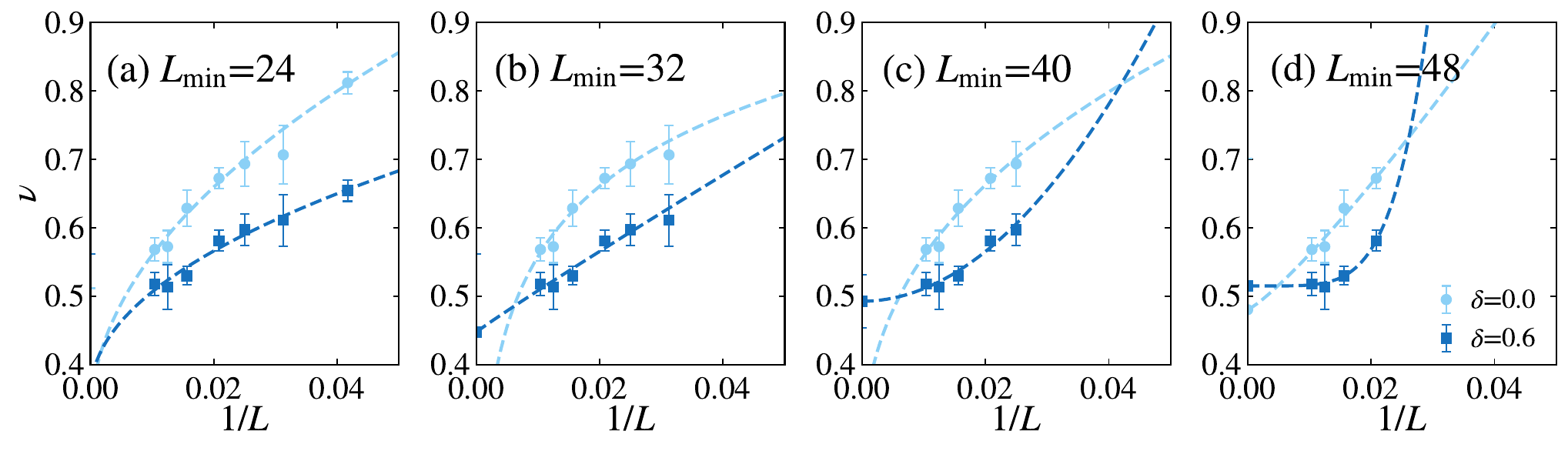}
\caption{ At $\Delta=0$, the critical exponent $\nu$ for the AFM order parameter are examined as functions of the system size at the critical transition point, under different strengths of non-Hermitian coupling: $\delta = 0.0$ and $ \delta = 0.6 $. 
The analysis utilizes fitting curves based on the power-law function as $\nu(L)=\nu_0+b/L^c$, with employing different minimum system size $L_{\text{min}}=24, 32, 40$ and 48. For the $L_{\text{min}}=48$, the fitting results are $\nu_0(\delta=0)\approx0.48$ and $\nu_0(\delta=0.6)\approx0.51$. }
\label{figs5}
\end{figure}
\section{Critical exponents}

The critical exponents $\eta(L)$ and $\nu(L)$ in Fig.\ref{fig4} of the main text at different system size are obtained based on structure factor and RG-invariant correlation ratio:
\begin{equation}\eta(L)=-\frac{1}{\log \left(2 \right)} \log (\frac{S_a(\vec{q}^{*}, 2L)}{S_a(\vec{q}^{*}, L)} )\vert_{g=g_{c}(L)}-(d+z-2)
\end{equation}
\begin{equation}1 / \nu(L)=\frac{1}{\log \left(2\right)} \log (\frac{\frac{\rm d}{{\rm d} g} R(2L)}{\frac{\rm d}{{\rm d} g} R(L)} )\vert_{g=g_{c}(L)}\end{equation}
\end{widetext}
\end{document}